\documentstyle[pre,aps,eqsecnum,epsf,graphicx]{revtex}
\begin{document}

\draft

\title{Model of multifragmentation, Equation of State and phase transition}
 
\author{C. B. Das$^1$, S. Das Gupta$^1$ and A. Z. Mekjian$^2$}

\address{$^1$Physics Department, McGill University, 
Montr{\'e}al, Canada H3A 2T8}
\address{$^2$Department of Physics and Astronomy, 
Rutgers University, Piscataway, New Jersey 08855}

\date{\today}

\maketitle

\begin{abstract}
We consider a soluble model of multifragmentation which is similar 
in spirit to many models which have been 
used to fit intermediate energy heavy ion collision data.
We draw a p-V diagram for the model and compare with a p-V diagram obtained
from a mean-field theory.  We investigate
the question of chemical instability in the multifragmentation model.
Phase transitions in the model are discussed.

\end{abstract}


\section{Introduction}
Statistical models of multifragmentation have long been used to
explain data from heavy ion collisons.  Such a model was first invoked
for Bevalac results \cite {Mekjian1,Dasgupta1} and similar physical ideas
but with many substantial variations were subsequently used for intermediate 
energy heavy ion collisions \cite {Bondorf,Koonin,Gross}.
In this work we consider a model of multifragmentation, variations of 
which have found many applications in the literature
\cite {Dasgupta2,Bhat1,Bhat2,Scott,Majumder,Dasgupta3,Tsang1}.  Thermodynamic
properties of a simpler version of this model have also been discussed
\cite {Bugaev1,Bugaev2,Elliott1}.  The model we use here has two kinds
of particles but no Coulomb interaction.  Throughout the rest of
this paper we will refer to this model as the thermodynamic model.
Typically the number of particles in our model is 200 although
we also use systems containing as many as 1000 particles.  While we could have
easily included a Coulomb interaction term, our objective here is different.
The aim here is to test if because of two kinds of particles two
features which have been discussed widely in recent literature (from
studies in mean field theory) persist in the thermodynamic model.
These features are chemical instability (analogous to mechanical
instability) and first order transition turning into second order. 
We therefore need to highlight some features which are present
both in the thermodynamic model and in 
mean-field theories when mean field theories are
applied to intermediate energy heavy ion physics.   Typically mean-field
theories use homogeneous infinite matter (hence no surface energy terms) 
and no Coulomb interaction.  Finite systems
with Coulomb and surface terms have also been included \cite{Lee} 
in mean-field 
models but this makes discussions more complicated and we want to 
stay at the simplest level.  As shown in \cite{Dasgupta2} surface energy
terms play an important role in a thermodynamic model and are included.
Moreover, since we will concentrate on two component systems, symmetry
energy terms are included as they are also in mean field theories.

\section{The Thermodynamic Model}

The thermodynamic model has been described in many places 
\cite{Dasgupta2,Bhat1,Dasgupta3}.  For completeness and to
enumerate the parameters we provide some details.

Assume that the system which breaks up after two ions hit each other
can be described as a hot equilibrated nuclear system characterized
by a temperature $T$ and a freeze-out volume $V$ within which there
are $A$ nucleons ($A=Z+N$).  The partition function of the system 
is given by
\begin{eqnarray}
Q_{Z,N}=\sum\prod_{i,j}\frac{\omega_{i,j}^{n_{i,j}}}{n_{i,j}!}
\end{eqnarray}
Here $n_{i,j}$ is the number of composites with proton number $i$
and neutron number $j$ and $\omega_{i,j}$ is the partition function
of a single composite with proton, neutron numbers $i,j$ respectively.
The sum is over all partitions of $Z, N$ into clusters and nucleons
subject to two constraints: $\sum_{i,j}in_{i,j}=Z$ and
$\sum_{i,j}jn_{i,j}=N$.  These constraints would appear to make
the computation of $Q_{Z,N}$ prohibitively difficult but a recursion
relation exists which allows the computation of $Q_{Z,N}$ quite easy
on the computer even for large $Z$ or $N$ \cite{C}.  Three equivalent recursion
relations exist, any one of which could be used.  For example, one
such relation is 
\begin{eqnarray}
Q_{z,n}=\frac{1}{z}\sum_{i,j}i\omega_{i,j}Q_{z-i,n-j}
\end{eqnarray}
The average number of particles of the species $i,j$ is given by
\begin{eqnarray}
\langle n_{i,j} \rangle=\omega_{i,j}\frac{Q_{Z-i,N-j}}{Q_{Z,N}}
\end{eqnarray}
All nuclear properties are contained in $\omega_{i,j}$.  It is given by
\begin{eqnarray}
\omega_{i,j}=\frac{V_f}{h^3}(2\pi (i+j)mT)^{3/2}\times q_{i,j,int}
\end{eqnarray}
Here $V_f$ is the free volume within which the particles move; $V_f$
is related to $V$ through $V_f=V-V_{ex}$ where $V_{ex}$ is the excluded 
volume due to finite sizes of composites.  This is the only interaction
between clusters we try to simulate. Thus the thermodynamic model is not 
an exact description of the  system considered here but another 
approximation to it which has some interesting
features that we hope to show. This restricts the validity of the
model to low density (i.e., large $V$).
Further, we take $V_{ex}$ to be
fixed, independent of multiplicity.  In reality, $V_{ex}$ should depend
upon multiplicity \cite{Majumder1}.  We take it to be constant and
equal to $V_0=A/\rho_0$ where $\rho_0$ is the normal nuclear density
and $A$ is the number of nucleons of the disassembling system.
As in previous applications, we restrict the model
to freeze-out densities less than $\rho/\rho_0=0.5$ that is $V\ge 2V_0$.  
The factor $q_{i,j,int}$
is the internal partition function of the composite.  Define $a=i+j$.
Then
\begin{eqnarray}
q_{i,j,int}=\exp{[Wa-\sigma a^{2/3}-s\frac{(i-j)^2}{a}+aT^2/\epsilon_0]/T}
\end{eqnarray}
Here $W=15.8$ MeV, $\sigma=18$ MeV, $s$=23.5 MeV and $\epsilon=16.0$ MeV.
The reader will recognise the volume, surface and symmetry energy of
the cluster $i,j$ and
the contribution to the internal partition function from excited states
in the fermi-gas formulation.  For $a(=i+j)\ge 5$ we use this formula.
For lower masses we simulate no Coulomb case by setting the binding
energy of $^3$He=binding energy of $^3$H and binding energy of 
$^4$Li=binding energy of $^4$H. In the weight of eq 2.4 we have not 
included a Fisher 
droplet model $\tau$ which is a power law prefactor that is important
around the critical point. Away from a critical point, exponential
terms dominant the weight and this is the region we study in this 
paper. Such a term can be included, but our main focus is on the role of
the symmetry energy in two component systems and related questions of
chemical instability.

For a given $a$, what are the limits on $i$(or $j=a-i$)? This is a 
non-trivial question.  In the results we will show,
we have taken limits by calculating the drip lines of protons and neutrons
as given by the above binding energy formula.  Limiting oneself
within the drip lines is a well-defined prescription, but is likely to be
an underestimation since resonances show up in particle-particle
correlation experiments.  On the other hand,
for a given $a$, taking the limits of $i$ from 0 to $a$ is definitely
an overestimation.   

There is another consideration which restricts the validity of the
model.  We have assumed (Eq.(2.1)) that the standard correction 
$n_{i,j}!$ takes
care of antisymmetry or symmetry of the particles.  In the range
$T>3$ MeV and $\rho/\rho_0<0.5$ this is usually true.  At low 
temperatures where one might apprehend the usual correction to fail,
it survives because many composites appear, thus there is not enough
of any particular species to make (anti)symmetrisation an important
issue.  At much higher temperature the number of protons and neutrons
increase but as is well-known, the $n!$ correction takes the approximate
partition function towards the proper one at high temperature.  We
define $y\equiv Z/(Z+N)$ where $Z$ and $N$ are the total proton
and neutron numbers of the disintegrating system
and the theory works even at low temperatures
if $y$ is in the vicinity of 0.5.  But for example, at $T$=5.0 MeV and
$y$=0 (neutron matter), this is a terrible model.  Now the number of
neutrons is large and the temperature is not high and Fermi-Dirac
statistics must be enforced.  This was studied quantitatively in
\cite{Jennings}.  In our applications of the thermodynamic model
we will confine $y$ to be between 0.3 to 0.7
and $T\ge $3 MeV.  This is indeed not very restrictive since this
encompasses the drip lines and so the model, which was devised for
intermediate energy heavy-ion collisions, will be applicable.  For
$^{124}$Sn+$^{124}$Sn collisions, a much studied case, the value of
$y$ is 0.4. 

By equation of state (EOS) we mean $p-V$ diagrams for fixed temperatures.
This can be obtained by exploiting the equation $p=T\frac{\partial lnQ_{Z,N}}
{\partial V_f}$.  From eqs (2.1) and (2.3), this reduces to 
$p=\frac{T}{V_f}(\sum \langle n_{i,j}\rangle-1)$ where -1 within the
parenthesis corrects for centre of mass motion.  The simplicity
of this formula suggests that we just have
a non-interacting gas with many species with $V$ being
replaced by $V_f$.  But this is deceptive.  In fact $\sum \langle n_{i,j}
\rangle$ (which is the multiplicity) is not fixed but varies as 
a function of both temperature $T$ and volume $V$ thus this is not anything
as simple as a mixture of non-interacting species.  It is indeed
interactions which make or break clusters to produce the final
equilibrium or statistical distribution of fragments.  In this sense
interactions are also included.

Since ours is a 
canonical model, we do not need the chemical potentials $\mu_p$ (proton
chemical potential) and $\mu_n$ but we compute them anyway
from the relation $\mu=(\frac{\partial F}{\partial n})_{V,T}$.  We
know the values of $Q_{Z,N}, Q_{Z-1,N}$ and $Q_{Z,N-1}$.  Since
$F$ is just $-Tln Q$, we compute $\mu_p$ from $\mu_p=-T(ln Q_{Z,N}-
ln Q_{Z-1,N})$ and $\mu_n$ from $-T(ln Q_{Z,N}-ln Q_{Z,N-1})$.  Indeed,
the grand canonical version of the thermodynamic model we are solving
has been known for a long time in heavy ion collision physics 
\cite{Dasgupta1}.  There the $\mu_p$ and $\mu_n$ arise naturally.
We have checked that the grand canonical values of $\mu_p$ and
$\mu_n$ are indeed very close to the ones we derive by exploiting
the canonical partition functions whose values we know numerically.
Throughout this work, whenever we plot $\mu$'s we have obtained
the values from a canonical calculation.  One might think that since
our model has many species there should be many $\mu$'s but
in fact all $\mu$'s can be expressed in terms of only $\mu_p$ and
$\mu_n$.  Since our model is based solely on phase space, chemical
equilibrium is in fact implied.

\section{A mean-field Model}
We want to contrast the model above with mean-field theories.
Our mean-field calculation uses the simplest model consistent with
nuclear matter binding energy, saturation density, compressibility
and symmetry energy for asymmetric matter.  The potential energy density 
is taken to be
\begin{eqnarray}
V(\rho_n,\rho_p)=\frac{A_u}{\rho_0}\rho_n\rho_p+\frac{A_l}{2\rho_0}
(\rho_n^2+\rho_p^2)+\frac{B}{\sigma+1}\frac{\rho^{\sigma+1}}{\rho_0^{\sigma}}
\end{eqnarray}
Here $\rho_0$=.16 fm$^{-3}$ and $\rho_n$ and $\rho_p$ are neutron and
proton densities and $\rho=\rho_n+\rho_p$.  The dimensionless constant $\sigma$
and $A_u, A_l$, $B$ (all in MeV) are chosen to reproduce nuclear
matter binding at 16 MeV per particle, saturation density at 0.16$fm^{-3}$,
compressibility at 201 MeV and symmetry energy at 23.5 MeV.
The energy per particle (including kinetic energy) at $T = 0$ is
\begin{eqnarray}
E/A=A_u\frac{\rho_n\rho_p}{\rho\rho_0}+\frac{A_l}{2\rho\rho_0}(\rho_n^2+
\rho_p^2)+\frac{B}{\sigma+1}(\frac{\rho}{\rho_0})^{\sigma}+
22.135\times [\frac{\rho_p}{\rho}(\frac{2\rho_p}{\rho_0)})^{2/3}+
\frac{\rho_n}{\rho}(\frac{2\rho_n}{\rho_0)})^{2/3}]
\end{eqnarray}
The values of the constants are: $\sigma=7/6; A_u=-379.2$ MeV;
$A_l=-334.4$ MeV; $B=$303.9 MeV.

The Hartree-Fock energy of an orbital is given by
\begin{eqnarray}
\epsilon=p^2/2m+A_u(\rho_u/\rho_0)+A_l(\rho_l/\rho_0)+B(\rho/\rho_0)^{\sigma}
\end{eqnarray}
The value of $\mu_p$ is found by solving for a given $\rho_p$ and
$\beta=1/T$
\begin{eqnarray}
\rho_p=\frac{8\pi}{h^3}\int_0^{\infty}\frac{p^2dp}
{\exp(\beta(\epsilon_p-\mu_p))+1}
\end{eqnarray}
Similarly $\mu_n$ is extracted from $\rho_n$.  The pressure has contributions
from kinetic energy and the potential energy.  The contribution from
the kinetic energy is calculated from well-known Fermi-gas model
formula.  Contribution to pressure from interaction is
\begin{eqnarray}
p_{skyrme}=\frac{A_u}{\rho_0}\rho_n\rho_p+
\frac{A_l}{2\rho_0}[(\rho_n)^2+(\rho_p)^2]+
\frac{\sigma}{\sigma+1}B(\frac{\rho}{\rho_0})^{\sigma}\rho
\end{eqnarray}

\section{EOS in the two models}

Fig. 1 compares the $p-\rho$ diagrams at constant temperature
for the two models.  We restrict the value of $y=\rho_p/(\rho_p+\rho_n)$
between 0.3 and 0.5 and $\rho/\rho_0$ between 0 and 0.5 because outside
these ranges the validity of the thermodynamic model is significantly reduced
although the mean-field model has no obvious limitations.  
In Fig.1 two curves for the mean-field model 
are shown for each temperature and $y$; one is the straightforward
$p-\rho$ diagram and the other is with Maxwell construction
which eliminates the spinodal instability region.
While the straightforward $p-\rho$ diagram in the mean field theory
is widely different from the the $p-\rho$ diagram of 
thermodynamic model,
the $p-\rho$ diagram with Maxwell construction is much closer
specially at temperature 7 MeV.  The $p-\rho$ diagrams at $T$=10 MeV
in the two models
are not that close.  At least part of the reason is that
the thermodynamic model
is drawn for exactly 200 nucleons but the mean-field theory uses
a grand canonical ensemble and hence is applicable to infinite
systems only.  We have checked that for a system of 500 nucleons
the thermodynamic $p-\rho$ diagram is closer to the Maxwell constructed
mean-field $p-\rho$ diagrams.

\vskip 0.2in
\epsfxsize=3.5in
\epsfysize=5.0in
\centerline{\rotatebox{270}{\epsffile{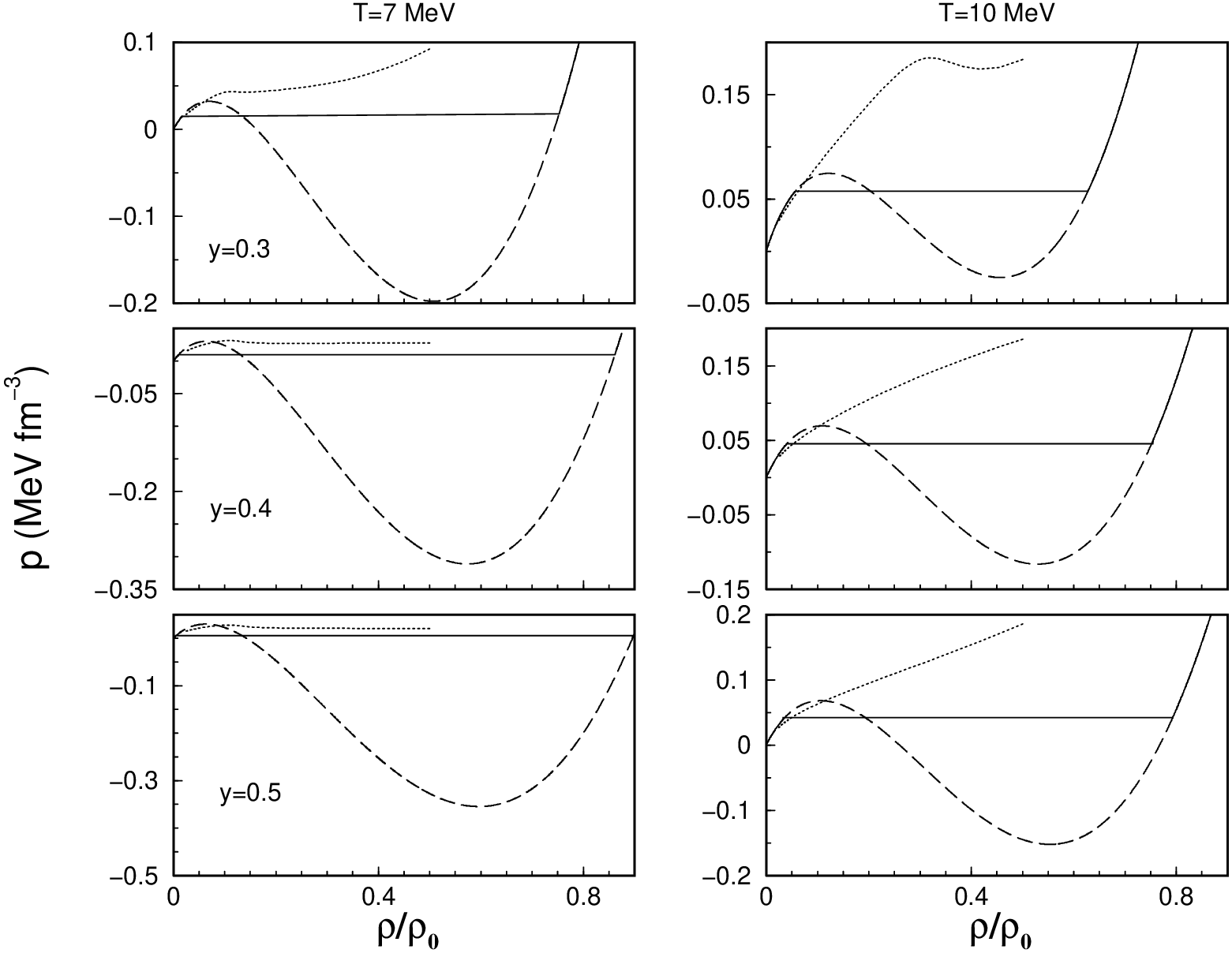}}}
\vskip 0.4in

\begin{center}
Fig.1: {\it The EOS in the two models at different temperatures 7
and 10 MeV (left and right panel respectively) and with different
proton fractions.  The dashed and dotted lines represent the EOS in
the meanfield and the thermodynamic model and the solid line is the
Maxwell construction in the meanfield model.}
\end{center}

We want to point out that regions of negative compressibility 
($dp/d\rho<0$) which are common in mean-field theory (Maxwell 
construction eliminates these) are almost absent in the 
thermodynamic model (they are present when plotted in an 
expanded scale, see Fig. 3) and one would be tempted to
conclude that the thermodynamic model is a good lowest order 
approximation. The thermodynamic model includes all inhomogeneous 
distributions of matter from single nucleons and light clusters 
with gas-like behavior to very large liquid-like clusters. 
This feature approximates the Maxwell construction incorporated 
into a mean field theory which splits the system
into two parts with liquid and gas densities. The very small region of
negative compressibilty left over has its origin probably in the
finite particle number effect and not is an inherent error in 
the model. This is dealt with again in section VIII.  
Mean field theory descriptions of two component systems introduce new 
features into the description not present in one component
systems. Specifically, a new variable has to be introduced, 
such as the proton fraction y, which can be different in the two 
phases. The liquid phase can have one value of y, with the gas 
phase having another value, while still maintaining the total 
number of protons and neutrons. With a new variable, the coexistence 
curve and instability curve of one component systems become 
surfaces in $p,T$, and $y$ for two components.
In mean field descriptions, the first order phase transition of one
component systems become a second order phase transition in two
component systems.  Moreover, mechanical instability and chemical 
instability no longer coincide. We now turn our attentions to features
associated with chemical instability in our model.

\section{isospin fractionation}
Isospin fractionation is a well-established experimental phenomenon
\cite{Xu}.  If the disintegrating system has a given $N/Z>1$ then,
after collision, the measured $n_n/n_p$ ratio (where $n_n, n_p$ are measured
single neutron and proton yields respectively) is higher than $N/Z$.
Similarly, the ratio of measured $<n_{1,2}>/<n_{2,1}>$ is higher than
what one might expect from the $N/Z$ ratio of the disintegrating system.
This then implies that if there is a large chunk left after the
breakup it must have a $n/z$ ratio lower than original $N/Z$ since the
total number of neutrons and protons must be conserved.  If we 
characterise $n/z$ ratio etc. in terms of the parameter $y$ we have
been using, then if $y_{source}$ is less than 0.5, then
$y$ of the large chunk is greater
than $y_{source}$ and $<n_p>/(<n_p>+<n_n>)$ is less than $y_{source}$.

{\it A priori}, it would seem difficult to get this aspect out of a
mean-field model but in a seminal piece of work Muller and Serot
have demonstrated how this might come about \cite{Muller1,Muller2}.
In mean-field theory,
analogous to mechanical instability ($\frac{\partial p}{\partial \rho}
<0$) there appears regions of chemical instability, i.e.,
$\frac{\partial \mu_p}{\partial y}<0$ (or $\frac{\partial \mu_n}
{\partial y}>0$) when two kinds of particles are involved.  One can
avoid this unphysical region of chemical instability 
but then needs to consider splitting the system into two
parts, each homogeneous but distinct from each other, 
one belonging to the liquid phase with higher $y$ value and the 
other to the gas phase with lower $y$ value.  One consequence
of this is that the phase transition takes place neither at constant
pressure nor at constant volume and what would have been a first order phase
transition, becomes a second order phase transition.

\vskip 0.2in
\epsfxsize=3.5in
\epsfysize=5.0in
\centerline{\rotatebox{270}{\epsffile{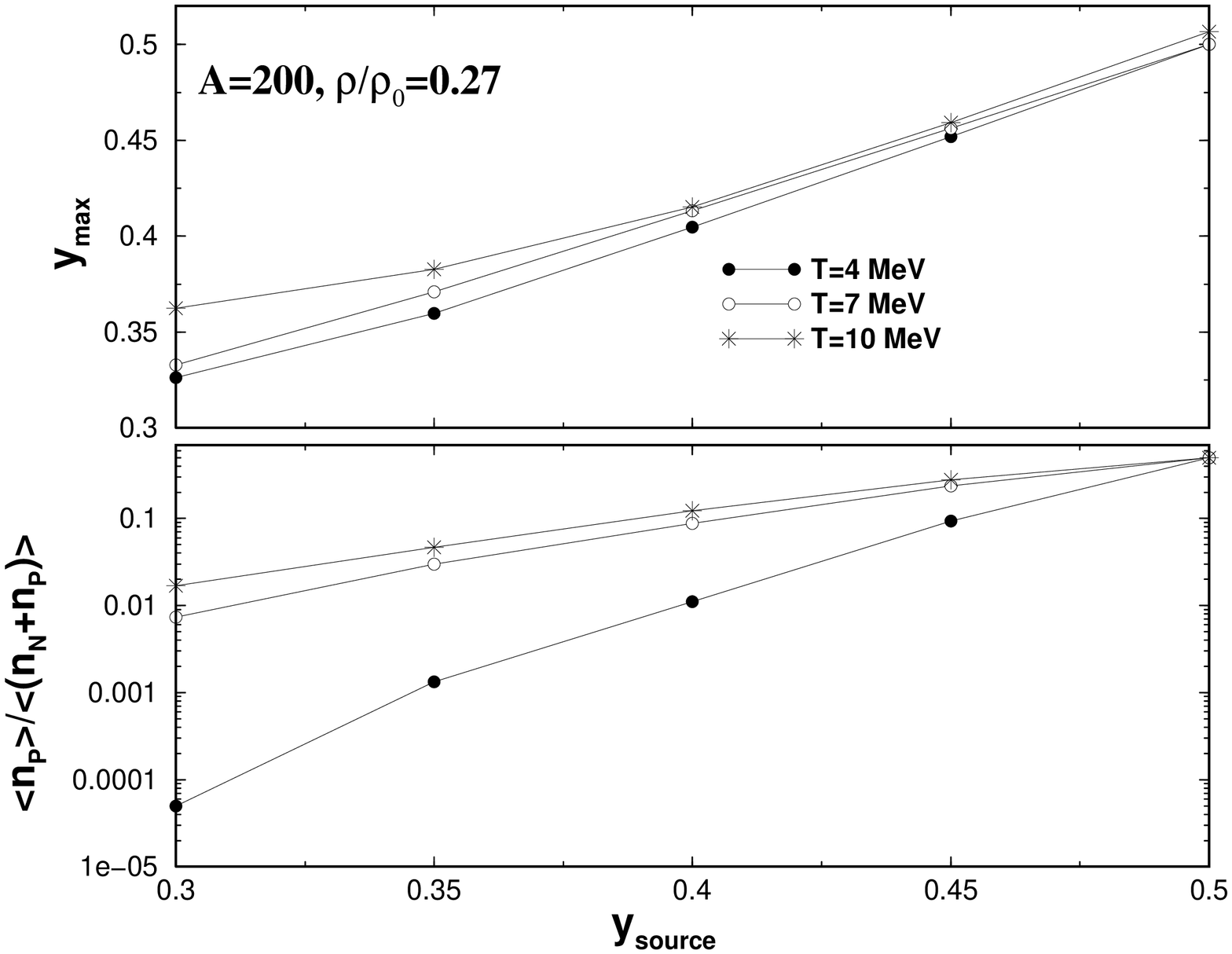}}}
\vskip 0.4in

\begin{center}
Fig.2: {\it Example of isospin fractionation in the thermodynamic model.   
$y$ of the largest cluster (top panel) is plotted in the top
panel.  This $y$ is larger than the $y$ value of the whole system.
The lower part of the figure shows that the gas of nucleons
is very rich.  For example for $y$=0.4 and $T$=7-10 MeV,
$\langle n_n\rangle\approx 10\langle n_p\rangle$ while for
$T$=4 MeV, $\langle n_n\rangle\approx 100\langle n_p\rangle$.}
\end{center}

In the thermodynamic model, isospin fractionation happens naturally.
In general, the model has, as final products, all allowed composites, 
$a,b,c,d$.... where the composite labelled $a$ has $y_a=i_a/(i_a+j_a)$ 
where $i_a, j_a$ is the number of protons and number of neutrons
respectively in the composite $a$.  The only law of conservation
is $Z=\sum_a i_a\times n_a$ and $N=\sum_a j_a\times n_a$.  So a large
chunk can exist with higher $y$ than that of the whole system and
populations of other species can adjust to obey overall
conservation laws.  Whatever partition lowers the free energy
will happen.  The thermodynamic model is dramatically different from
mean field models.  The most significant difference is that
in the thermodynamic model, if we
prescribe that dissociation takes place at $\rho/\rho_0$=0.3 we
still have only clusters with normal nuclear density and
properties and also nucleons.  It is just that there are empty 
spaces between different
clusters and nucleons in the region of dissociation.  But in mean field models
$\rho/\rho_0$=0.3 will imply that the nuclear matter is uniformly
stretched to this density.  While this can happen as a transient 
phenomenon such as in transport calculation, whether this can also
exist as an equilibrium situation is highly questionable.

An example of isospin fractionation in the thermodynamic model is shown
in Fig. 2.  The formalism developed in section II can also be extended
to calculate the average number of nucleons and protons (or neutrons)
in the largest cluster.  For brevity we do not write down the formulae
here but these are straightforward extensions of Eqs.(2.7) and (2.8)
given in \cite{Dasgupta2}.  Fig. 2 shows results from such a calculation.
If $y$ of the disintegrating system
is less than 0.5, the $y$ value of the largest cluster is larger than
that of the source.  Correspondingly, $(\langle n_p\rangle)/
(\langle n_p\rangle +\langle n_n\rangle)$ is much smaller that $y_{source}$.
[It should be mentioned that the number of protons and neutrons will
be augmented from decays of hot composites, so what is plotted in
Fig. 2 is not
what will actually be observed in experiments].  Further the isospin
fractionation happens whether the dissociation takes
place at constant volume or constant pressure.

In this and many other aspects, the thermodynamic model is very similar
to the Lattice Gas Model (LGM) with isospin dependence.  For an accurate
solution of LGM one has to give up the mean-field approach and
obtain results by Monte-Carlo simulation.  Here also
many composites are produced with many different $y$ values
\cite{Pan,Chomaz}.  Isospin fractionation happens naturally \cite{Chomaz}.

\section{Instability in the thermodynamic model}
Fig. 1 shows that compared to the mean-field model, regions of 
mechanical instability
with $\frac{\partial p}{\partial \rho}<0$ nearly disappear in the
thermodynamic model.  In an expanded scale, they are more readily
seen (Fig. 3) where we have drawn $p-\rho$ diagram for a constant 
temperature $T$=7.0 MeV but different $y$'s.

\vskip 0.2in
\epsfxsize=3.5in
\epsfysize=5.0in
\centerline{\rotatebox{270}{\epsffile{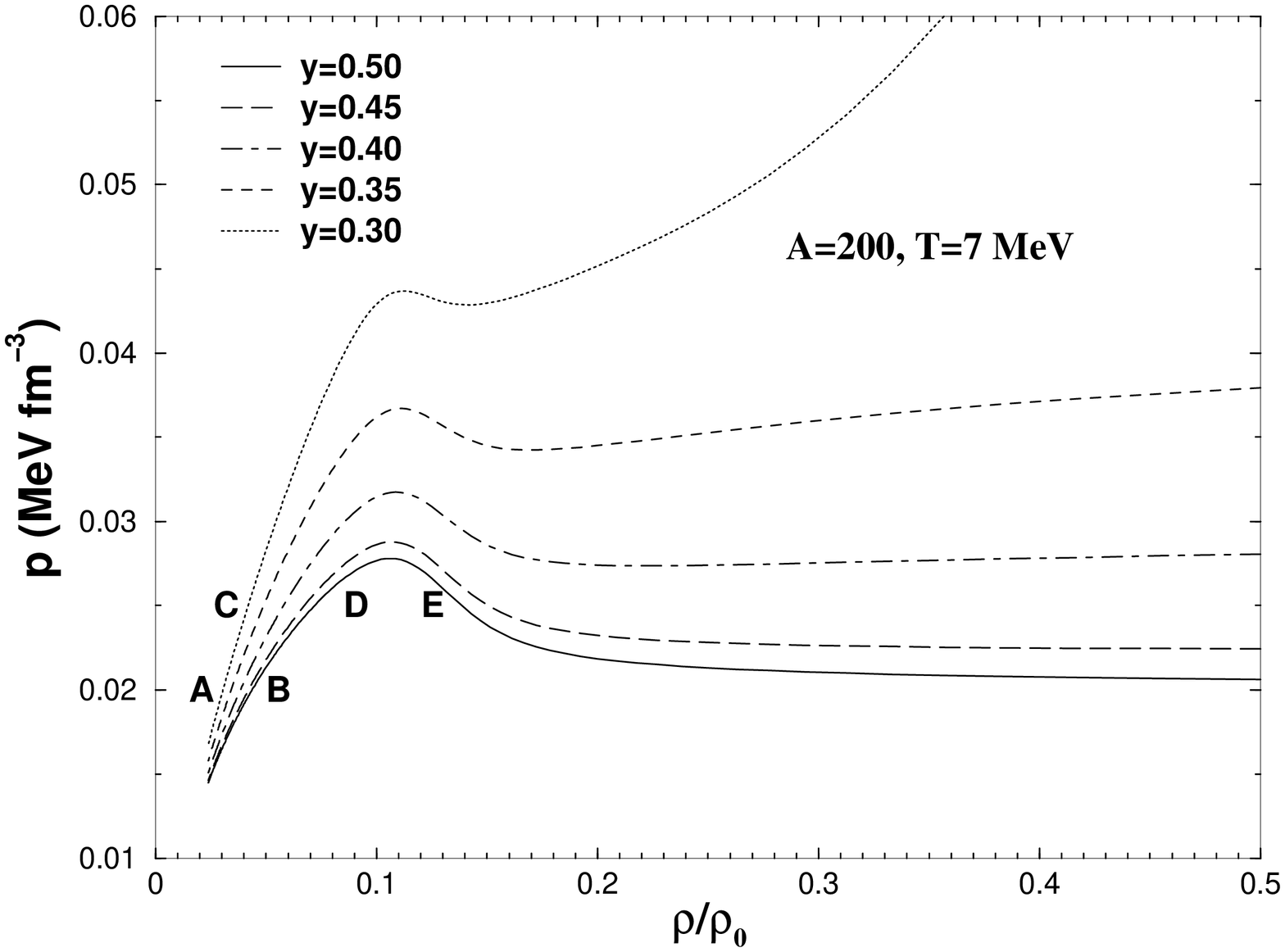}}}
\vskip 0.4in

\begin{center}
Fig.3: {\it EOS ($p-\rho$) in the thermodynamic model at $T$=7.0 MeV
but for different $y$'s.}
\end{center}

We clearly have some regions of mechanical instability.  Chemical 
instability implies $(\frac{\partial \mu_p}{\partial y})_{p,T}< 0$.
We investigate that now.
At $T$=7 MeV, we have drawn $\mu_p$ (and $\mu_n$) at four 
pressures (Fig. 4).  To get an understanding of the behaviour,
we need to also look at Fig. 3.
At the lowest pressure shown, $p=0.02$ MeV $fm^{-3}$ (Fig. 4), the 
horizontal constant pressure curve cuts the isothermals (Fig. 3) at the low
density side only (between $A$ and $B$) and $\mu_p$ rises monotonically
between $y=0.3$ and $y=0.5$.  The next constant pressure curve, at
$p=0.025$ MeV $fm^{-3}$ (Fig. 4) cuts all isothermals 
(Fig. 3) at low density side 
($\rho/\rho_0< 0.1$) between $C$ and $D$ and a few isothermals at
higher density side.  Between $C$ and $D$, $y$ increases as does $\mu_p$.
The points marked $D$ and $E$ have the same
values of $p$ and $T$ but very slightly differing values of $\mu_p$.
As we move to the right from $E$ along the line $p$=0.025 MeV $fm^{-3}$
the value of $y$ drops as also the value of $\mu_p$.  We forego describing
graphs at other pressures but the figure shows 
there is a very small region where $\frac{\partial \mu_p}{\partial \rho}$
is negative ( Fig. 4, $p=0.035$ MeV $fm^{-3}$).  The not so obvious
feature is the appearence of two branches 
in both $\mu_p$ and $\mu_n$ (i.e., for example, the $p=
0.025$ MeV $fm^{-3}$ curve).  The two branches would merge for a 
Van der Waals fluid and thus the appearence of two branches indicates
departure from a Van der Waals fluid.

\section{Comparison with Van der Waals fluid}
For a Van der Waals fluid with Maxwell construction,
the following behaviour will be seen \cite{Reif} as we move along
an isothermal in a $\mu-\rho$ plane,
provided we are below the critical temperature.
If we start with very small density we are in the gas phase.
As the density rises, the chemical potential changes till it
reaches the coexistence region.  In this region $\mu_{gas}=\mu_{liq}$
and as the density changes the chemical potential remains unchanged 
but more particles change from the gas phase to the liquid phase.
This remains the situation till a high density is reached when all
the particles are in the liquid phase.

\vskip 0.2in
\epsfxsize=3.2in
\epsfysize=4.5in
\centerline{\rotatebox{270}{\epsffile{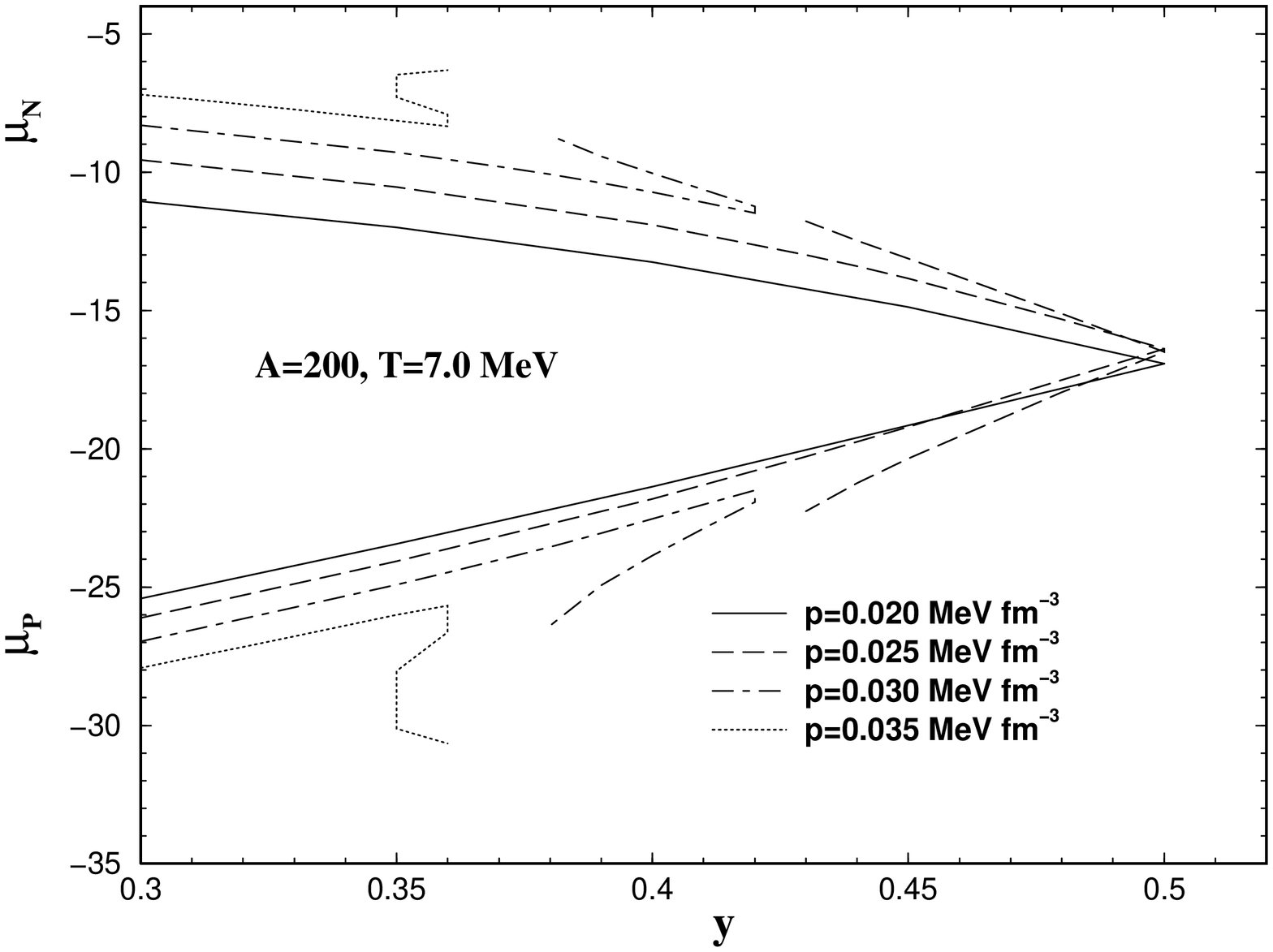}}}
\vskip 0.2in

\begin{center}
Fig.4: {\it $\mu_P$ and $\mu_N$ as a function of $y$. 
Here $T$=7 MeV.}
\end{center}

\vskip 0.2in
\epsfxsize=3.5in
\epsfysize=4.5in
\centerline{\rotatebox{270}{\epsffile{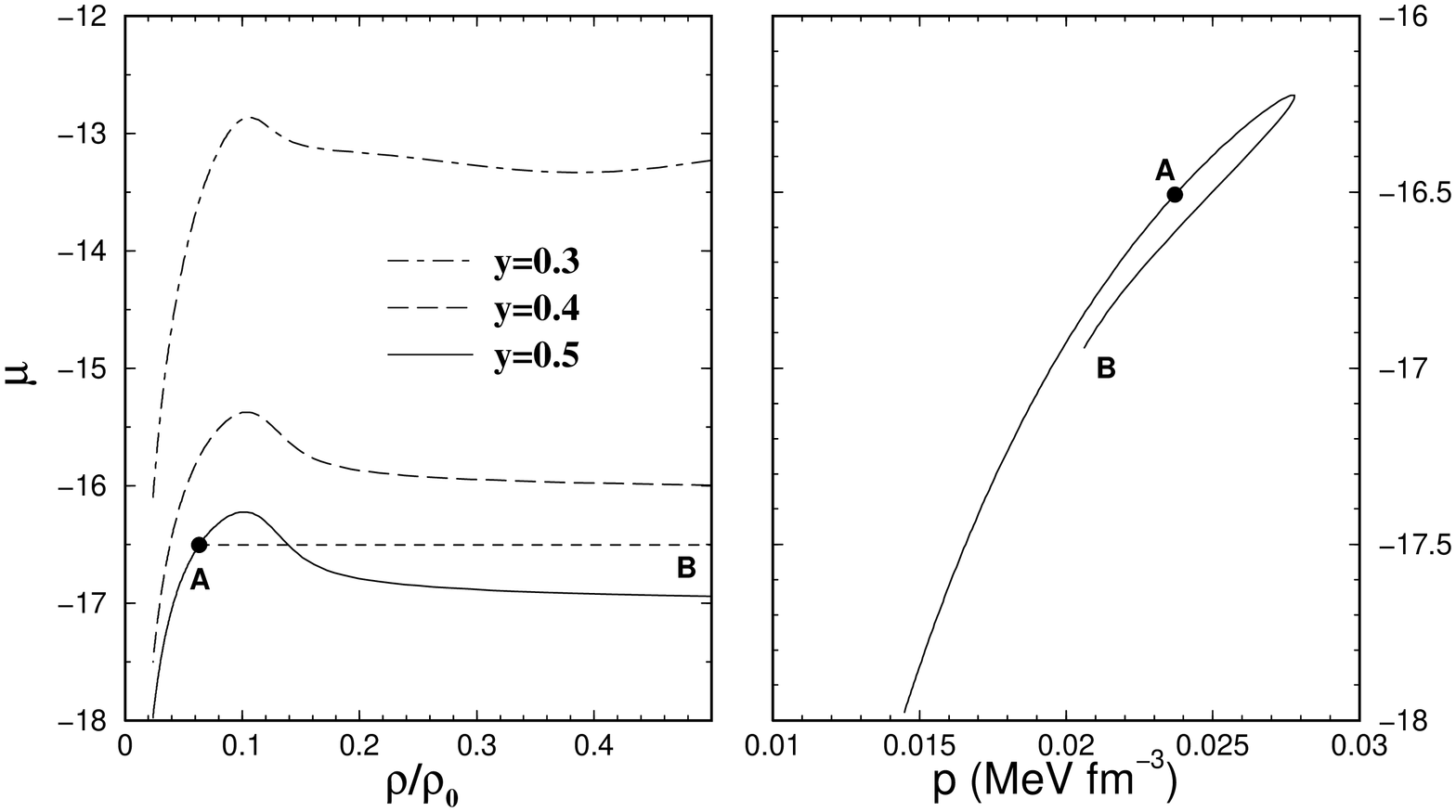}}}
\vskip 0.2in

\begin{center}
Fig.5: {\it $\mu=y\mu_p+(1-y)\mu_n$ as a function of $\rho/\rho_0$
for different $y$'s(left panel).  The temperature is 7 MeV.
In the right panel the behaviour of $\mu$ is shown as a function
of pressure $p$ for $y=0.5$ and $T=$7 MeV.}
\end{center}

The situation in the thermodynamic model is depicted in Fig. 5.
The model is inapplicable to high density situation, so we cut-off
$\rho$ at $\rho/\rho_0=0.5$.  We have two chemical potentials
$\mu_p$ and $\mu_n$ and to compare to a Van der Waals fluid, we
consider the combination $\mu=y\mu_p+(1-y)\mu_n$
(for $y$=0.5, $\mu_p=\mu_n$ any way).  At $T$=7.0 MeV we show
in the left panel of Fig.5 the behaviour of $\mu$ at $y=0.5, 0.4$
and 0.3.  For the $y$=0.5 curve we also depict schematically what the
behaviour would have been if we had a Maxwell-corrected 
Van der Waals fluid.
From some point $A$, the chemical potential would remain unchanged
(shown by a horizontal line ending at $B$ which is the end-point
of our density).  A more familiar plot is $\mu$ against pressure $p$
for a fixed temperature.  This is shown in the right panel for our
model.  For a Van der Waals fluid, the segment from $A$ to $B$ would
simply collapse to the point $A$.

\section{specific heats in the model}

In \cite{Dasgupta2} where the thermodynamic model was first studied
for phase transitions, it was pointed out that for a given density
$\rho$, the specific heat per particle $C_V/A$ tends to $\infty$ at a 
particular temperature when the particle number $A$ tends to $\infty$.
Since $C_V=(\frac{\partial E}{\partial T})_V=T(\frac{\partial S}{\partial T})_V
=-T(\frac{\partial^2F}{\partial^2T})_V$, a singularity in $C_V$ signifies
a break in the first derivative of $F$, the free energy and a first
order phase transition.  The model in
\cite{Dasgupta2} considered one kind of particle
although binding energy, surface energy etc. were chosen
to mimic the nuclear case.  We see similar effect here when we 
take into account two kinds of particles explicitly (Fig. 6, see
also \cite{Bhat1}).  The
calculated $C_V/A$ becomes progressively sharply peaked as $A$ 
increases for all $y$ values between 0.3 and 0.5.  This behaviour
of the specific heat is very different from that of mean field model
of nuclear matter where the specific heat at constant volume varies
smoothly from a low temperature Fermi gas to an ideal gas as $T$ increases.
In a thermodynamic model with fragmentation, this behaviour is modified
by the surface energies that arise in the multifragmentation of the
original nucleus into clusters of different sizes.  The peak in the 
specific heat occurs at the point where the largest cluster
suddenly disappears.  This behaviour is nuclear boiling.

Specific heat per particle $C_p/A$ in the model has not been considered
before.  We notice there are regions with $\frac{\partial p}{\partial\rho}<0$
(even though these regions are much less visible than in the mean
field model).  Their presence might indicate finite particle number effects
or it may be a shortcoming of the model.  Calculations which are on
going suggest it is particle number effect rather than an inherent
problem in the model.  These negative regions of compressibility
can lead to negative values of $C_p$.  This is a contentious issue
at the moment and we intend to deal with these issues fully in a 
future publication.

\vskip 0.2in
\epsfxsize=3.5in
\epsfysize=5.0in
\centerline{\rotatebox{270}{\epsffile{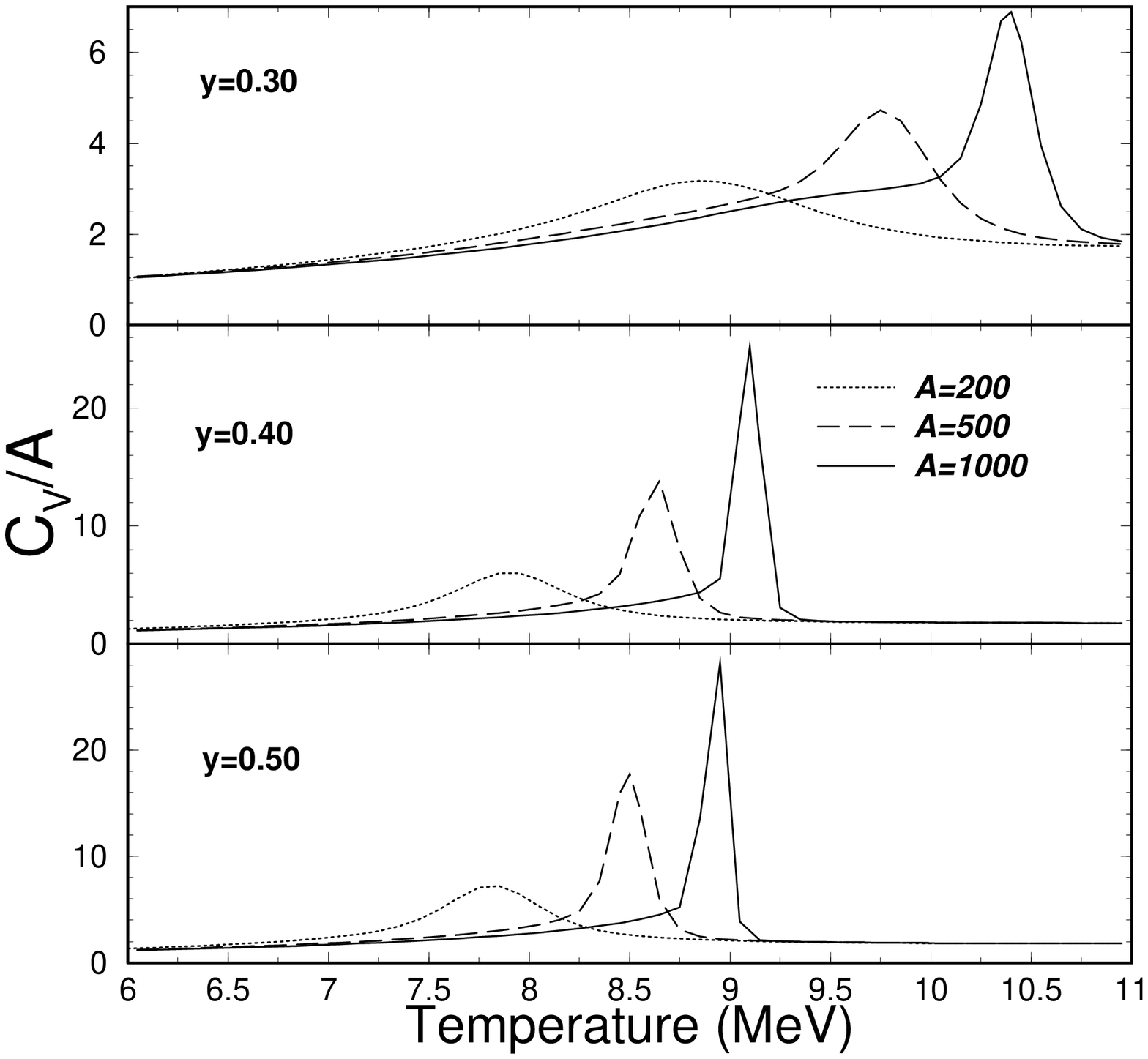}}}
\vskip 0.3in

\begin{center}
Fig.6: {\it $C_V/A$ as a function of temperature for systems of 200, 500
and 1000 particles with different proton fractions.}
\end{center}

\section{Summary}
We looked at several features of a thermodynamic model
(which has seen many applications in data fitting) and a mean field
model.  Equation of state in mean-field theory has large regions
of mechanical instability even for infinite systems and one
needs to do a Maxwell construction to eliminate these.  
By contrast, the thermodynamic model has directly an EOS
which becomes very flat with density and volume and this behaviour
resembles a real system undergoing a first order phase transition.
In mean-field models, when the system enters the region of
instability, it fragments into pieces.  This fragmentation is
directly included in the thermodynamic model and this is the reason
for relative flatness in the EOS.  The cluster distribution
readjusts itself with changes in $V$ or $\rho$ to maintain a nearly
constant pressure.
Isospin fractionation seen in experiments can
be also obtained in the mean field model but it requires a
bifurcation in the isotopic space.  It also requires that during
dissociation neither pressure nor volume remain constant.  By
contrast, isospin fractionation occurs naturally in the thermodynamic
model and can happen either at constant volume or at constant pressure.
Large differences between these two models also appear in the calculation
of $C_V$.  The thermodynamic model has a strong peak in $C_V$ whose
origin are surface energy terms in the multifragmentation process
which is lacking in a mean field model of homogeneous nuclear matter.
This peak is associated with the phenomenon of nuclear boiling and
the sudden disappearence of the largest cluster.

\section{Acknowledgment}
This work is supported in part by the Natural Sciences and Engineering
Research Council of Canada and the U.S. Department of Energy Grant No. 
DE FG02-96ER40987.

\end{document}